\documentclass[pra,aps,floats,twocolumn]{revtex4}
\usepackage{graphicx}
\usepackage{amsmath}

\begin{document}
\title{Fractionalized long-range ordered state in a Falicov-Kimball model
}
\author{Minh-Tien Tran}
\affiliation{Institute of Research and Development, Duy Tan University, K7/25 Quang Trung, Danang, Vietnam  \\
Institute of Physics, Vietnam Academy of Science and Technology,
Hanoi 100000, Vietnam}

\begin{abstract}
A Falicov-Kimball model which thermodynamically reduces the local Coulomb interaction of particles to attraction or repulsion is studied within the dynamical mean-field theory. In the strong interaction regime a fractionalization of particles into charge and spin objects, the physical properties of which are different from the whole particles, is observed in both high- and low-temperature phases. At high temperature and strong interaction the single-particle density of states opens an excitation gap, but the charge compressibility and the spin susceptibility exhibit the features of gapless excitations. The low-temperature phase has a long-range order, and the single-particle spectra are always gapped, while the charge and spin excitations are gapless in the strong interaction regime. In the fractionalized long-range ordered phase both the charge compressibility and the spin susceptibility are universal scaling functions of temperature.
\end{abstract}


\maketitle

\section{Introduction}

In the many-body physics the fractionalization is the phenomenon where the particles of the system can be constructed as
combinations of objects with new
quantum numbers. The physical properties of the system cannot be determined by combinations of its elementary constituents. One of the prominent examples is the one-dimensional system of interacting electrons \cite{Giamarchi}. In one dimension the electrons are fractionalized into charge and spin objects, and the low-energy properties of the system are determined by collective excitations of these charge and spin objects \cite{Giamarchi}.
Other example may include the so-called orthogonal metal, which has recently attracted research attention \cite{Senthil}. It is a non-Fermi liquid, in which the transport and thermodynamics are like the Fermi-liquid ones, but the quasiparticle is absent.
The fractionalization is an intriguing effect of strong electron correlations.
It is not only fascinating in itself, but has also been suggested to be the key element in understanding the nature of different phenomena such as the Mott insulator or high-temperature superconductors. The Mott insulator can be interpreted as a quantum spin liquid, where its $1/2$ spin quasiparticles do not carry charge \cite{Ng1,Ng2}. The normal state of high-temperature superconductors exhibits unusual metallic properties, which seem to be related to a non-Fermi liquid \cite{Schofield}.

Recently, Hohenadler and Assaad have introduced a Falicov-Kimball model (FKM), which thermodynamically reduces the local Coulomb interaction to attraction or repulsion \cite{Assaad}. This FKM can be considered as a three-component generation of the standard
spinless FKM \cite{Tran1,Tran2,Tran3}. The spinless FKM, or alternatively, a simplified Hubbard model, where one of the two spin species is movable, and the other is localized,  was introduced as a minimal model for studying various phenomena such as a semiconductor-metal transition, crystallization, and correlations in alloys \cite{Hubbard1,Hubbard2,Gutzwiller,FK,Lieb,Gruber}.
The presence of localized fermions leads the metallic state, which occurs at weak correlations to be non-Fermi liquid \cite{FK} or an Anderson localization \cite{Antipov}. At low temperature the FKM exhibits different exotic ordered states \cite{Freericks,Lemanski}. The FKM has attracted research attention due to its rich physics and its simplicity compared with the Hubbard model.
Quantum Monte-Carlo simulations, which are performed for the FKM proposed by Hohenadler and Assaad on a two-dimensional square lattice, reveal an exotic metal in the strong correlation regime \cite{Assaad}. In the exotic metallic phase the single-particle spectra are gapped, but the charge and spin excitations are gapless \cite{Assaad}. This demonstrates while the charge and spin excitations are like the metallic ones, the quasiparticle is absent. The exotic metal is indeed a fractionalized state. The FKM proposed by Hohenadler and Assaad is rarely a minimal lattice model among more sophisticated ones, which can exhibit an electron fractionalization \cite{Kitaev,Fu}. So far, the electron fractionalization is only realized in a metallic state without any long-range order.

In this work, we show the electron fractionalization can also coexist with a long-range order. In this electron fractionalization the single-particle spectra still open a gap. However, the gap opening is due to a long-range ordering. Despite the gap opening, the charge compressibility and the spin susceptibility exhibit the gapless excitation features. The opposite behaviours of electrons and their charge and spin counterparts lead the long-range ordered phase to be fractionalized.  We will
show this realization of the electron fractionalization in the FKM proposed by Hohenadler and Assaad at low-temperature. Actually, the FKM proposed
by Hohenadler and Assaad is a special symmetric case of the generalized three-component FKM with a three-body interaction \cite{Tran1,Tran2,Tran3}.
The three-component FKM exhibits various Mott insulators with different natures \cite{Tran1,Tran2,Tran3}.
In contrast to the previous studies \cite{Assaad,Tran1,Tran2,Tran3}, in this work we focus on the low-temperature phase, where a long-range ordering may occur. We use the dynamical mean-field theory (DMFT) to investigate a possibility of electron fractionalization. The DMFT is a widely and successfully used tool for treating strong electron correlations in a self-consistent non-perturbative manner\cite{Metzner,GKKR}. It is exact in the infinite dimensional limit \cite{Metzner}. The FKM was also successfully solved by the DMFT \cite{Mielsch1,Mielsch2,Mielsch3,FZ}.
The DMFT solutions of the FKM capture the essential features of the exact solutions at finite dimensions.
Within the DMFT,
at low temperature and strong correlation regime, we find a charge (or spin) long-range ordered state, in which the single-particle spectra are gapped, while the charge compressibility and the spin susceptibility are like metallic ones. In the fractionalized state the charge compressibility and the spin susceptibility obey a universal scaling law. In addition, the DMFT also allows us to study both the high-temperature metal-insulator transition (MIT) and low-temperature ordering in detail.
Within the DMFT we could calculate the single-site double and triple occupancies, which are accessible by the site-resolved imaging techniques \cite{Sherson,Bakr,Greif}. This gives a possibility of comparing the theoretical results with experiments.
With advantages of ultracold techniques the proposed FKM can be realized in an optical lattice, and this could verify the electron fractionalization
in the proposed model.

The present paper is organized as follows. In Sec. II we present the model and its DMFT. The numerical results are presented in Sec. III.
Finally, the conclusion is presented in Sec. IV .

\section{Model and dynamical mean-field theory}
We study the FKM proposed by Hohenadler and Assaad for a fractionalized metallic state \cite{Assaad}.
The proposed FKM describes a lattice of two-component movable and single-component localized particles.
Its Hamiltonian reads
\begin{eqnarray}
H=-t\sum_{\langle ij\rangle,\sigma} ( c^\dagger_{i\sigma} c_{j\sigma} + \text{H.c.}) +
U \sum_{i} Q_i \prod_\sigma \big(n_{i\sigma}-\frac{1}{2}\big),
\label{ham1}
\end{eqnarray}
where $c^{\dagger}_{i\sigma}$ ($c_{i\sigma}$) is the
creation (annihilation) operator of a conduction electron with spin $\sigma$ at lattice site $i$.
$n_{i\sigma} =  c^\dagger_{i\sigma} c_{i\sigma}$ is the number operator. $t$ is the hopping parameter between the nearest-neighbour sites.
The localized fermions are present in the model through their Ising degree of freedom
$Q_i = \pm 1$. $U$ is a three-body interaction, which is a combination of
the Hubbard interaction
of conduction electrons and the Ising variable. When $Q_i=\pm 1$ the three-body interaction is reduced to the repulsive (attractive)
Hubbard interaction of conduction electrons.
Hamiltonian in Eq.~(\ref{ham1}) is a special case of the three-component FKM \cite{Tran1,Tran2,Tran3} with a three-body interaction
\begin{eqnarray}
H &= & -t\sum_{\langle ij \rangle,\sigma} ( c^{\dagger}_{i\sigma} c_{j\sigma} + \text{H.c.}) +
E_Q \sum_{i} Q_{i}  \nonumber \\
&& +
V \sum_{i\sigma} Q_i n_{i\sigma}
+ U \sum_{i} Q_i n_{i\uparrow} n_{i\downarrow} , \label{ham2}
\end{eqnarray}
where $E_Q$ is the energy level of localized spinless fermions, $V$ is the Falicov-Kimball interaction between conduction electrons and localized fermions, and
$U$ is their three-body interaction. The Ising variable $Q_i$  is connected to the localized spinless fermions in the three-component FKM via the relation
$Q_i = 2 n^{\text{loc}}_i -1$, where $n^{\text{loc}}_i$ is the number operator of the localized spinless fermions \cite{Tran1,Tran2,Tran3}.
Although both the Ising variable $Q_i$ and the number operator $n^{\text{loc}}_i$ are equivalent, since they are conserved, the FKM written in the terms of the Ising variable omits the explicit dynamics of the localized fermions. In the FKM, the dynamics of the localized fermions is non-trivial \cite{FZ,Brandt}. Actually,
the Ising variable $Q_i$ was also previously introduced that reduces the Falicov-Kimball interaction into a staggered magnetic field in a proposal of the FKM \cite{Lieb}.
However, the three-body interaction in Eq.~(\ref{ham2}) already contains the local two-body interaction
$-U \sum_{i} n_{i\uparrow} n_{i\downarrow}$ of conduction electrons when the Ising variable is replaced by its number operator counterpart.
The explicit correlations of conduction electrons distinguish the three-component FKM from the spin extension of the FKM, where the local interaction between conduction electrons is absent \cite{Lieb,FZ}.
In this work we focus on the dynamics of conduction electrons.

When $E_Q=U/4$ and $V=-U/2$, Hamiltonians in Eqs.~(\ref{ham1})-(\ref{ham2}) are identical.
Hamiltonian in Eq.~(\ref{ham1}) or in Eq.~(\ref{ham2}) can be realized by loading ultracold atoms in an optical lattice. Actually, Hamiltonian in
Eq.~(\ref{ham1}) is the Hubbard model with randomly alternating local interactions. The standard Hubbard model has already been realized by quantum simulations of ultracold atoms \cite{Jordens,Schneider}. A spatial modulation of the local interaction has also been achieved \cite{Yamazaki}. This leads to a possibility of realizing the Hubbard model with spatially alternating local interactions by quantum simulations \cite{Koga}.
Hamiltonian in Eq.~(\ref{ham2}) can also be simulated by loading two-component light and single-component heavy fermionic atoms, for instance
$^6$Li and $^{173}$Yb, into an optical lattice. In a sufficient deep lattice, the heavy atoms can be localized, and only the light atoms are movable through the lattice. The three-body and few-body interactions have also been achieved in ultracold atoms \cite{Will}. With a symmetric tuning of the model parameters, Hamiltonian in Eq.~(\ref{ham1}) can also be realized through the three-component FKM.

We consider a bipartite lattice, which can be divided into two penetrating sublattices $A$ and $B$. The single particle properties of conduction electrons can be determined by their Green function
\begin{equation}
\mathbf{G}_\sigma(\mathbf{k},z) = \langle\langle \Psi_{\mathbf{k}\sigma} | \Psi_{\mathbf{k}\sigma}^\dagger \rangle\rangle_z ,
\label{green}
\end{equation}
where $\Psi_{\mathbf{k}\sigma}^\dagger = ( a^\dagger_{\mathbf{k}\sigma} ; b^\dagger_{\mathbf{k}\sigma} )$, and $a^\dagger_{\mathbf{k}\sigma}$,
$ b^\dagger_{\mathbf{k}\sigma}$ are the creation operators for conduction electrons in the sublattice $A$ and $B$, respectively.
In Eq.~(\ref{green}) we have used the Zubarev's notation for the double-time Green function \cite{Zubarev}. The Green function $\mathbf{G}_\sigma(\mathbf{k},z)$ is actually the Fourier transform of the
retarded (or advanced) Green function in the time domain \cite{Zubarev}
\begin{eqnarray*}
\mathbf{G}_\sigma(\mathbf{k},t) = \mp i \theta(\pm t) \langle \{\Psi_{\mathbf{k}\sigma}(t), \Psi_{\mathbf{k}\sigma}^\dagger\} \rangle,
\end{eqnarray*}
where $\theta(t)$ is the Heaviside step function, and the curly brakets denote the anticommutator $\{A,B\}\equiv AB + BA$.
We will use the DMFT to solve the FKM described in Eq. (\ref{ham2}) with $E_Q=U/4$ and $V=-U/2$ at half filling
in a similar way of the solving three-component FKM \cite{Tran1}.
Within the DMFT, the self energy of conduction electrons is a local function of frequency. From the Dyson equation we obtain
\begin{equation}
\mathbf{G}_\sigma(\mathbf{k},z) = \left(
\begin{array}{cc}
z +\mu - \Sigma_{A\sigma}(z) & -\varepsilon_{\mathbf{k}} \\
-\varepsilon_{\mathbf{k}} & z +\mu -\Sigma_{B\sigma}(z)
\end{array} \right)^{-1} ,
\end{equation}
where $\Sigma_{\alpha\sigma}(z)$ is the self energy of conduction electrons in the sublattice $\alpha$ ($\alpha=A, B$), and $\varepsilon_{\mathbf{k}}$ is the dispersion of conduction electrons, and $\mu$ is the chemical potential. The self energy is determined from a single correlated site embedded in an effective medium. The action of the embedded single site of the $\alpha$ sublattice reads
\begin{eqnarray}
\mathcal{S}_{\alpha} &=& \int_{0}^{\beta} d\tau \big(\sum_{\sigma}
\Psi^\dagger_{\alpha\sigma}(\tau) [-\mathcal{G}_{\alpha\sigma}(\tau)]^{-1} \Psi_{\alpha\sigma}(\tau) + E_Q Q_\alpha \nonumber \\
&&
 + V \sum_{\sigma} Q_\alpha n_{\alpha\sigma}(\tau) + U Q_\alpha n_{\alpha\uparrow}(\tau) n_{\alpha\downarrow}(\tau)
\big), \label{ac}
\end{eqnarray}
where the Green function $\mathcal{G}_{\alpha\sigma}(\tau)$ represents the effective mean-field medium, which contains all correlation effects of whole lattice except for the considered site in a mean-field manner. It can be determined by the Dyson equation
\begin{equation}
\mathcal{G}_{\alpha\sigma}^{-1}(z) = G_{\alpha\sigma}^{-1}(z) + \Sigma_{\alpha\sigma}(z),
\end{equation}
where $G_{\alpha\sigma}(z)$ is the local Green function of conduction electrons in the sublattice $\alpha$. We consider the hypercubic lattice in infinite dimensions. The local Green function is calculated by
\begin{equation}
G_{\alpha\sigma}(z) = \int d\varepsilon \rho_{0}(\varepsilon) [\mathbf{G}_{\sigma}(\varepsilon,z)]_{\alpha\alpha} ,
\end{equation}
where
\begin{equation*}
\rho_{0}(\varepsilon) = \frac{1}{t^{*} \sqrt{\pi}} \exp(-\varepsilon^2 / t^{*2}) ,
\end{equation*}
is the density of states (DOS) of noninteracting conduction electrons in the infinite-dimensional hypercubic lattice \cite{Metzner,GKKR}. $t^*$ is the rescaling hopping parameter in the infinite-dimensional limit. We will use $t^*$ as the energy unit.

Since $Q_\alpha$ is a good quantum number, we can take the trace over it in calculating the partition function of the single site problem
\begin{eqnarray}
\mathcal{Z}_\alpha &=& \text{Tr}_{Q_\alpha} \int \mathcal{D}[\Psi^\dagger_\sigma,\Psi_\sigma] \exp[-\mathcal{S}_\alpha] \nonumber \\
&=& \sum_{l=\pm 1} \exp(-\beta l E_{Q}) \mathcal{Z}_{\alpha l},
\end{eqnarray}
where
\begin{eqnarray}
\mathcal{Z}_{\alpha l} &=& \int \mathcal{D}[\Psi^\dagger_\sigma,\Psi_\sigma] \exp[-\mathcal{S}_{\alpha l}] ,  \\
\mathcal{S}_{\alpha l} &=& \int_{0}^{\beta} d\tau \big(\sum_{\sigma}
\Psi^\dagger_{\alpha\sigma}(\tau) [-\mathcal{G}_{\alpha\sigma}(\tau)]^{-1} \Psi_{\alpha\sigma}(\tau) \nonumber \\
&&
+ l V \sum_{\sigma}  n_{\alpha\sigma}(\tau) + l U  n_{\alpha\uparrow}(\tau) n_{\alpha\downarrow}(\tau)
\big). \label{action}
\end{eqnarray}
$\mathcal{S}_{\alpha l}$ in Eq.~(\ref{action}) is actually the action of an effective single site of the Hubbard model with the local interaction
$l U$ and the chemical potential shifted by $l V$. It gives the local Green function
\begin{eqnarray}
g_{\alpha l \sigma}(z) = \frac{1}{\mathcal{G}^{-1}_{\alpha\sigma}(z) - l V - \Xi_{\alpha l \sigma}(z)} ,
\end{eqnarray}
where $\Xi_{\alpha l \sigma}(z)$ is the self energy due to the local Hubbard interaction $l U$.
The local Green function of the original effective single site described by the action in Eq.~(\ref{ac}) is
\begin{eqnarray}
G_{\alpha \sigma}(z) = \sum_{l=\pm 1} w_{\alpha l} g_{\alpha l \sigma}(z) , \label{loc}
\end{eqnarray}
where
\begin{eqnarray}
w_{\alpha l} = \frac{\mathcal{Z}_{\alpha l} \exp(-\beta l E_Q)}{\mathcal{Z}_{\alpha}} .
\end{eqnarray}
Equation (\ref{loc}) shows the local Green function $G_{\alpha \sigma}(z)$ contains electron correlations which are generated from
both the repulsive ($l=1$) and attractive ($l=-1$) interactions. $w_{\alpha l}$ represents the weight factor of the contributions of the repulsive ($l=1$) or attractive ($l=-1$) interactions to the dynamics of conduction electrons. Within the DMFT we are able to explicitly study the contributions of the repulsive or attractive interactions to the dynamics of the system.  One can show that
\begin{eqnarray}
\langle Q_\alpha \rangle = \frac{1}{\mathcal{Z}_\alpha} \frac{\partial \mathcal{Z}_\alpha}{\partial E_Q}
= w_{\alpha,l=1} - w_{\alpha,l=-1} .
\end{eqnarray}
This shows the expectation value $\langle Q_\alpha \rangle$ measures the difference of the weight factors of
the repulsive and attractive interactions in the system. When $\langle Q_\alpha \rangle=0$ both the repulsive and attractive interactions
equally contribute to the Green function.
$\langle Q_\alpha \rangle=\pm 1$ indicates only the repulsive (or attractive) interaction plays the dominant role.

We calculate the self energy $\Xi_{\alpha l \sigma}(z)$ of the action in Eq.~(\ref{action}) by the exact diagonalization \cite{GKKR,Krauth}.
Within the exact diagonalization procedure the action $\mathcal{S}_{\alpha l}$ is mapped into an Anderson impurity model
\begin{eqnarray}
\lefteqn{
H_{\alpha l} = (l V - \mu) \sum\limits_{\sigma} c^\dagger_{\alpha\sigma} c_{\alpha\sigma} + l U n_{\alpha\uparrow} n_{\alpha\downarrow} }\nonumber \\
& + \sum\limits_{m\sigma} E_{\alpha m\sigma} d^\dagger_{\alpha m\sigma} d_{\alpha m\sigma} + \sum\limits_{m\sigma} V_{\alpha m \sigma} c^\dagger_{\alpha\sigma} d_{\alpha m\sigma} +\text{H.c.} , \label{and}
\end{eqnarray}
where the creation and annihilation operators $d^\dagger_{\alpha m\sigma}$, $d_{\alpha m\sigma}$ represent a finite set of $N_s$ orbitals $m$
which are a discrete mapping of the effective medium. The parameters $E_{\alpha m\sigma}$, $V_{\alpha m \sigma}$ are determined by the minimization of the mapping difference of the effective medium Green function in the Matsubara frequency domain \cite{GKKR,Krauth}. With a finite orbital set, Hamiltonian in Eq.~(\ref{and}) can exactly be diagonalized, and we are able to calculate the local Green function $g_{\alpha l \sigma}(z)$ from the Lehmann spectral representation \cite{GKKR,Krauth}.
Following the iteration procedure of the DMFT \cite{GKKR,Krauth}, we could obtain the local Green function $G_{\alpha \sigma}(z)$ and the self energy
$\Sigma_{\alpha\sigma}(z)$ self consistently.

Once the self-consistent solution is achieved, we compute the charge compressibility and the spin susceptibility. The charge compressibility
is defined as
\begin{equation}
\kappa = \frac{1}{n^2} \frac{1}{N} \sum_{i\sigma} \frac{\partial \langle n_{i\sigma} \rangle}{\partial \mu},
\end{equation}
where $n=\sum_{i\sigma} \langle n_{i\sigma} \rangle /N$, and $N$ is the number of lattice sites. In order to compute the spin susceptibility, we introduce an external magnetic field $h_i$ , which applies to conduction electrons
\begin{equation}
H_{\text{mf}} = \frac{1}{2} \sum_{i\sigma} h_{i} c^\dagger_{i\sigma} \sigma c_{i\sigma}.
\end{equation}
We consider both uniform $h_i=h$, and staggered $h_i=(-1)^i h$ magnetic fields. In the case of uniform magnetic field the spin susceptibility is a ferromagnetic (FM) one
\begin{equation}
\chi_{\text{FM}} = \frac{1}{2N} \sum_{i\sigma} \sigma \frac{\partial  \langle  n_{i\sigma} \rangle  }{\partial h}\bigg|_{h=0} ,
\end{equation}
and in the case of staggered magnetic field, the spin susceptibility is an antiferromagnetic (AF) one
\begin{equation}
\chi_{\text{AF}} = \frac{1}{2N} \sum_{i\sigma} (-1)^i \sigma \frac{\partial \langle  n_{i\sigma} \rangle  }{\partial h}\bigg|_{h=0} .
\end{equation}
We use the Ridder implementation of numerical derivatives to calculate the charge compressibility and the spin susceptibility in the numerical calculations \cite{NR}.

\section{Numerical results}

\begin{figure}[t]
\begin{center}
\includegraphics[width=0.48\textwidth]{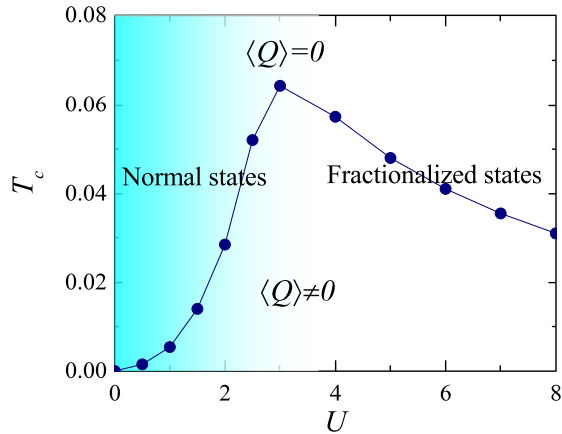}
\end{center}
\caption{(Color online) The critical temperature $T_c$ via the local interaction $U$. The region $T > T_c$ is the HM phase, while the
region $T<T_c$ is the long-range ordered phase. As $U$ increases, the phase continuously changes from the normal to the fractionalized states at both high and low temperatures.}
\label{fig1}
\end{figure}

We numerically solve the set of the DMFT equations by iterations. The effective single impurity problem is solved by the exact diagonalization \cite{GKKR,Krauth}.
In numerical calculations we typically use $N_s=4$ orbitals. We have also checked the results with $N_s=5$. We mainly focus the study on
the half filling case, in which $\mu=0$.
At high temperature we obtain $\langle Q_\alpha \rangle =0$, while at low temperature $\langle Q_\alpha \rangle \neq 0$.
The high temperature solution $\langle Q_\alpha \rangle =0$ leads to a homogeneous (HM) phase, where $\langle n_{i\sigma} \rangle = 1/2$.
At low temperature we obtain different long-range ordered solutions: charge ordered (CO) and AF phases.
The solution $\langle Q_\alpha \rangle < 0$ is accompanied by the CO phase, while $\langle Q_\alpha \rangle > 0$ appears together with the AF phase.
The CO phase is paramagnetic and it
is characterized by staggered electron density
$\langle n_{i\sigma} \rangle = n/2 + (-1)^i \Delta_{\text{CO}}$, where $\Delta_{\text{CO}}$ is the order parameter of the charge ordering.
In the AF phase
the electron density obeys $\langle n_{i\sigma} \rangle = n/2 + (-1)^i \sigma \Delta_{\text{AF}}$, where $\Delta_{\text{AF}}$ is
the order parameter of the AF ordering. At half filling $n=1$ although the symmetry between the $A$ and $B$ sublattices is broken due to the long-range ordering,
the sublattice symmetric solution $\langle Q_A \rangle = \langle Q_B \rangle$ is still maintained for the Ising variable. Away from half filling, a solution $\langle Q_A \rangle \neq \langle Q_B \rangle$ is obtained at low temperature.
The low temperature solutions are obtained depending on the initial input self energy. An initial CO (AF) self energy leads to the CO (AF) solution.
The CO and AF phases appear below the same critical temperature $T_c$. In Fig.~\ref{fig1} we plot the critical temperature as a function of $U$. The obtained $T_c$ qualitatively agrees with the Monte-Carlo simulation result \cite{Assaad}, despite it is the infinite-dimensional result.
The critical temperature approaches to zero in both the limits of weak and strong interactions.
The HM phase was previously studied by the Monte-Carlo simulation in Ref.~\onlinecite{Assaad}. However, there is a lack of studies on the long-range ordered phases. In addition, within the DMFT the explicit contributions of the repulsive and attractive interactions to the system dynamics are calculable in both the HM and the long-range ordered phases.

\subsection{Homogeneous phase}

\begin{figure}[t]
\begin{center}
\includegraphics[width=0.4\textwidth]{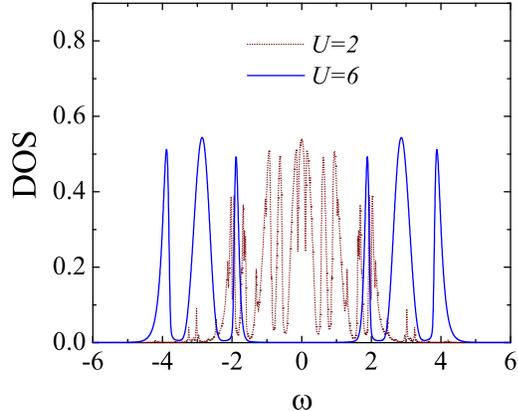}
\end{center}
\caption{(Color online) The DOS of conduction electrons in the HM phase for different values of $U$ at temperature $T=0.1$.}
\label{fig2a}
\end{figure}

The DMFT results of the HM phase agree well with the reported results of the Monte-Carlo simulation \cite{Assaad}. Indeed, the HM phase is separated into two regimes. One is the weak correlation regime, where the DOS shows the metallic behavior. The other is the strong correlation regime, where the single-particle DOS opens a gap. In Fig.~\ref{fig2a} we plot the DOS in both the weak and strong correlation regimes. We have used the Lorentzian broadening parameter $\eta=0.01$ for the delta functions in the DOS.  The gap opening in the strong correlation regime indicates the single particle excitation is similar to the one in an insulator. In the present model, the local DOS is a linear combination of the repulsive ($U>0$) and attractive ($U<0$) interaction DOS, as it is shown in Eq.~(\ref{loc}). At half filling ($\mu=0$) and $V=-U/2$, the shift of the chemical potential $\pm V$ in the effective single-impurity problem in Eq.~(\ref{action}) keeps the cases of repulsive and attractive interactions always on half filling. At half filling, the repulsive and the attractive Hubbard models are equivalent due to the particle-hole symmetry in the bipartite lattice \cite{Toschi,Keller}. Therefore, at low temperature the metallic and insulating solutions may coexist \cite{GKKR,Georges,Rozenberg}. However, this coexistence occurs below the critical temperature of the long-range ordering, and these coexistent solutions are unstable in respect to the long-range ordered phase.

\begin{figure}[t]
\begin{center}
\includegraphics[width=0.4\textwidth]{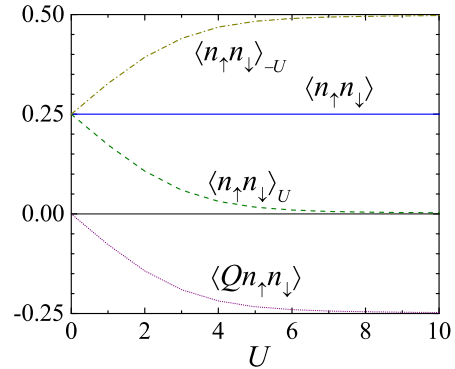}
\end{center}
\caption{(Color online) The double $\langle n_{\uparrow} n_{\downarrow} \rangle$ (solid line) and triple
$\langle Q n_{\uparrow} n_{\downarrow} \rangle$ (shot dotted line) occupancies via interaction $U$ in the HM phase at temperature $T=0.5$. The contributions of the repulsive $\langle n_{\uparrow} n_{\downarrow} \rangle_{U}$ (dashed line) and attractive $\langle n_{\uparrow} n_{\downarrow} \rangle_{-U}$ (dash dotted line) interactions to the double occupancy are also shown.}
\label{fig2c}
\end{figure}

Within the DMFT we can calculate local quantities such as the double $\langle n_{\uparrow} n_{\downarrow} \rangle$ and triple $\langle Q n_{\uparrow} n_{\downarrow} \rangle$ occupancies from the effective single-site problem.
The double occupancy is used to experimentally detect the MIT in optical lattices \cite{Jordens,Schneider}. It can also be obtained from the site occupation, which is accessible by the site-resolved imaging experiments \cite{Sherson,Bakr,Greif}.
From the action of the effective single site in Eq.~(\ref{ac}), one can show
\begin{eqnarray}
\langle n_{\uparrow} n_{\downarrow} \rangle = w_{1} \langle n_{\uparrow} n_{\downarrow} \rangle_{U} + w_{-1} \langle n_{\uparrow} n_{\downarrow} \rangle_{-U} , \\
\langle Q n_{\uparrow} n_{\downarrow} \rangle = w_{1} \langle n_{\uparrow} n_{\downarrow} \rangle_{U} - w_{-1} \langle n_{\uparrow} n_{\downarrow} \rangle_{-U} ,
\end{eqnarray}
where $\langle n_{\uparrow} n_{\downarrow} \rangle_{\pm U}$ is the double occupancy in the repulsive (attractive) Hubbard action described by Eq.~(\ref{action}) with $l=\pm 1$.
In the strong correlation regime the repulsive Hubbard interaction suppresses the double occupancy
$\langle n_{\uparrow} n_{\downarrow} \rangle_{U>0} \rightarrow 0$,  while the attractive one binds the local pair of electrons with opposite spins, hence
$\langle n_{\uparrow} n_{\downarrow} \rangle_{U<0} \rightarrow n/2$. \cite{Keller}
Since in the HM phase $w_{1}=w_{-1}=1/2$, thus $\langle n_{\uparrow} n_{\downarrow} \rangle \rightarrow n/4$, and
$\langle Q n_{\uparrow} n_{\downarrow} \rangle \rightarrow -n/4$ in the strong correlation limit.
In Fig.~\ref{fig2c} we plot the double and triple occupancies as functions of the interaction.
One can imagine these occupancies as a bonding of the corresponding quantities in the Mott insulator ($U>0$) and in the electron pairing state ($U<0$).
Accidentally, the double occupancy
in the HM phase at half filling $n=1$ is independent on the interaction strength $U$ as it is shown in Fig.~\ref{fig2c}. However, the double occupancy in the weak and the strong correlation limits has different origins. In the limit $U \rightarrow 0$,
$\langle n_{\uparrow} n_{\downarrow} \rangle = (n/2)^2$, while in the limit $U \rightarrow \infty$,
$\langle n_{\uparrow} n_{\downarrow} \rangle = n/4$ is a strong correlation effect.
The smooth dependencies of the double and triple occupancies on the interaction suggest that the correlation-driven MIT in the HM phase is just a continuous crossover from metal to insulator. With the advantages of the site resolved imaging technique, the double and triple occupancies would be measured as functions of the interaction, once the present model is simulated by ultracold atoms.

\begin{figure}[t]
\includegraphics[width=0.48\textwidth]{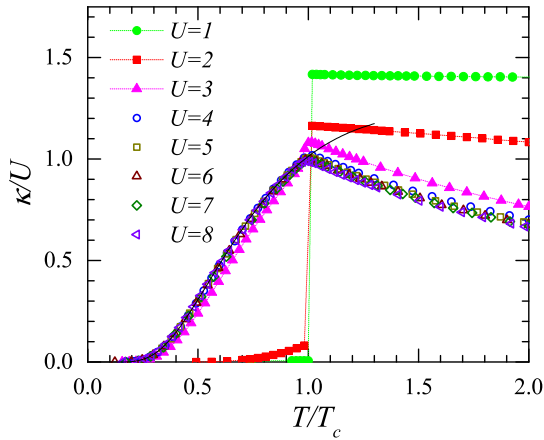}
\caption{(Color online) Scaling of the charge compressibility $\kappa/U$ as a function of $T/T_c$.
The black solid line is the fitting function, which is described in Eq. (\ref{fitting}) with $a=7.6$, $b=1.945$, and $c=0.25$.}
\label{fig3}
\end{figure}

\begin{figure}[b]
\includegraphics[width=0.5\textwidth]{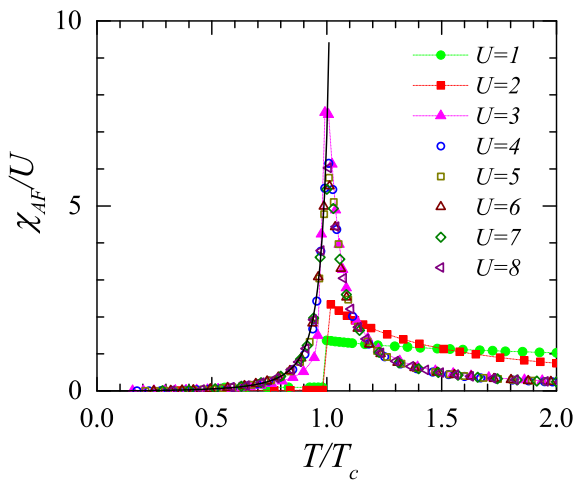}
\caption{(Color online) Scaling of the spin susceptibility $\chi_{\text{AF}}/U$ as a function of $T/T_c$. The black solid line
is the fitting function, which is described in Eq. (\ref{fitting}) with $a=0.0175$, $b=0.568$, and $c=-1.698$.}
\label{fig4}
\end{figure}

\begin{figure}[t]
\begin{center}
\includegraphics[width=0.4\textwidth]{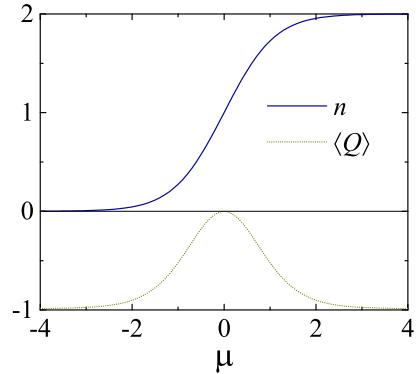}
\end{center}
\caption{(Color online) The electron density $n$ and $\langle Q \rangle$ via the chemical potential $\mu$ in the strong correlation regime $U=5$ at temperature $T=0.5$.}
\label{fig2d}
\end{figure}

In the strong correlation regime, the single-particle DOS opens a gap, which is usually an insulator's attribute. However, the charge and spin excitations show non-insulating behaviors. In Figs.~\ref{fig3} and \ref{fig4} we plot the  charge compressibility and the AF spin susceptibility. We obtain the FM spin susceptibility $\chi_{\text{FM}}=\kappa/4$. The HM phase occurs at $T > T_c$. In the HM phase the charge compressibility and the spin susceptibility are always finite for any finite interaction $U$. This indicates the charge and spin excitations are gapless, and their behaviors are qualitatively the same for both the weak and strong correlation regimes. However, the single-particle excitation in the strong correlation regime are gapped. The opposite behaviors of the single-particle excitation and its charge and spin counterparts  constitute the HM phase a fractionalized state in the strong correlation regime. In this fractionalized state the single-particle properties look like the insulating ones, but the charge compressibility and the spin susceptibility exhibit the metallic feature. The finite value of the charge compressibility can also be seen from the dependence of the electron density on the chemical potential. In Fig.~\ref{fig2d} we plot the electron density $n$ and $\langle Q \rangle $ as functions of the chemical potential in the strong correlation regime. The electron density monotonously increases with the chemical potential, hence the charge compressibility is finite. However, only at half filling $\langle Q \rangle =0$. Away from half filling $\langle Q \rangle $ is finite, and the phase becomes ordered.
The disordered state $\langle Q \rangle =0$ is just unique in the surrounding of ordered phases.
When the chemical potential is shifted from its half-filling value $\mu=0$, the state nature is also changed.
Therefore, the charge compressibility does not vanish despite the gap opening in the single-particle spectra at half filling.

\subsection{Long-range ordered phase}

\begin{figure}[b]
\begin{center}
\includegraphics[width=0.35\textwidth]{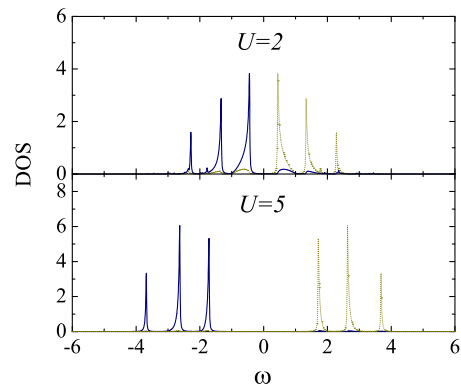}
\end{center}
\caption{(Color online)The DOS of conduction electrons in the charge ordered phase at temperature $T=0.01$. The solid (dotted) lines are the DOS in the sublattice $A$ ($B$).}
\label{fig6}
\end{figure}

\begin{figure}[t]
\begin{center}
\includegraphics[width=0.45\textwidth]{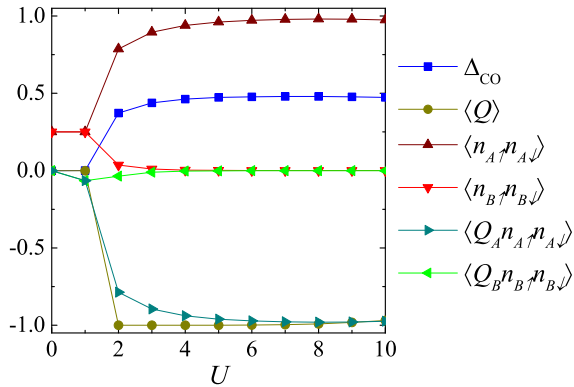}
\end{center}
\caption{(Color online) The charge order parameter $\Delta_{\text{CO}}$, $\langle Q \rangle$, the double and triple occupancies via the interaction $U$ at temperature $T=0.01$.}
\label{fig7}
\end{figure}

Below the critical temperature a long-range ordering occurs. Both conduction electrons and the Ising variable are ordered. Although at low temperature we obtain two solutions, the CO phase with $\langle Q \rangle <0 $ and the AF  phase with $\langle Q \rangle >0$, their charge compressibility and spin susceptibility are the same for both phases. Therefore, we focus the present study on the CO phase. The single-particle DOS in the CO phase is always gapped. In Fig.~\ref{fig6} we plot the DOS for both the weak and strong correlation regimes. The gap opening due to the long-range ordering indicates the single-particle excitation is similar to the one in a Slater insulator. In Fig.~\ref{fig7} we plot the charge order parameter $\Delta_{\text{CO}}$, $\langle Q \rangle$, as well as the double and triple occupancies as functions of the interaction $U$. In the strong correlation regime, one sublattice, for instance $A$, is fully occupied, while the other sublattice ($B$) is empty. In the empty sublattice the double and triple occupancies vanish. In the occupied sublattice, the attractive interaction gives the dominant contributions to the double and triple occupancies, hence $\langle n_{A\uparrow}n_{A\downarrow} \rangle \rightarrow n_A/2=1$ when $U \gg 1$. In addition,  in the CO phase
$w_{1} \rightarrow 0$, and $w_{-1} \rightarrow 1$, which result in $\langle Q \rangle \rightarrow -1$. The CO phase is also a pairing state, where pairs of electrons with opposite spins are bound at every sites of the occupied sublattice due to the attractive interaction. Figure \ref{fig7} also shows the phase transition from the HM phase ($\langle Q \rangle =0 $) to the CO one ($\langle Q \rangle \neq 0 $) when the interaction $U$ increases. It is consistent with the phase diagram plotted in Fig. \ref{fig1} at a fixed temperature.

\begin{figure}[t]
\begin{center}
\includegraphics[width=0.4\textwidth]{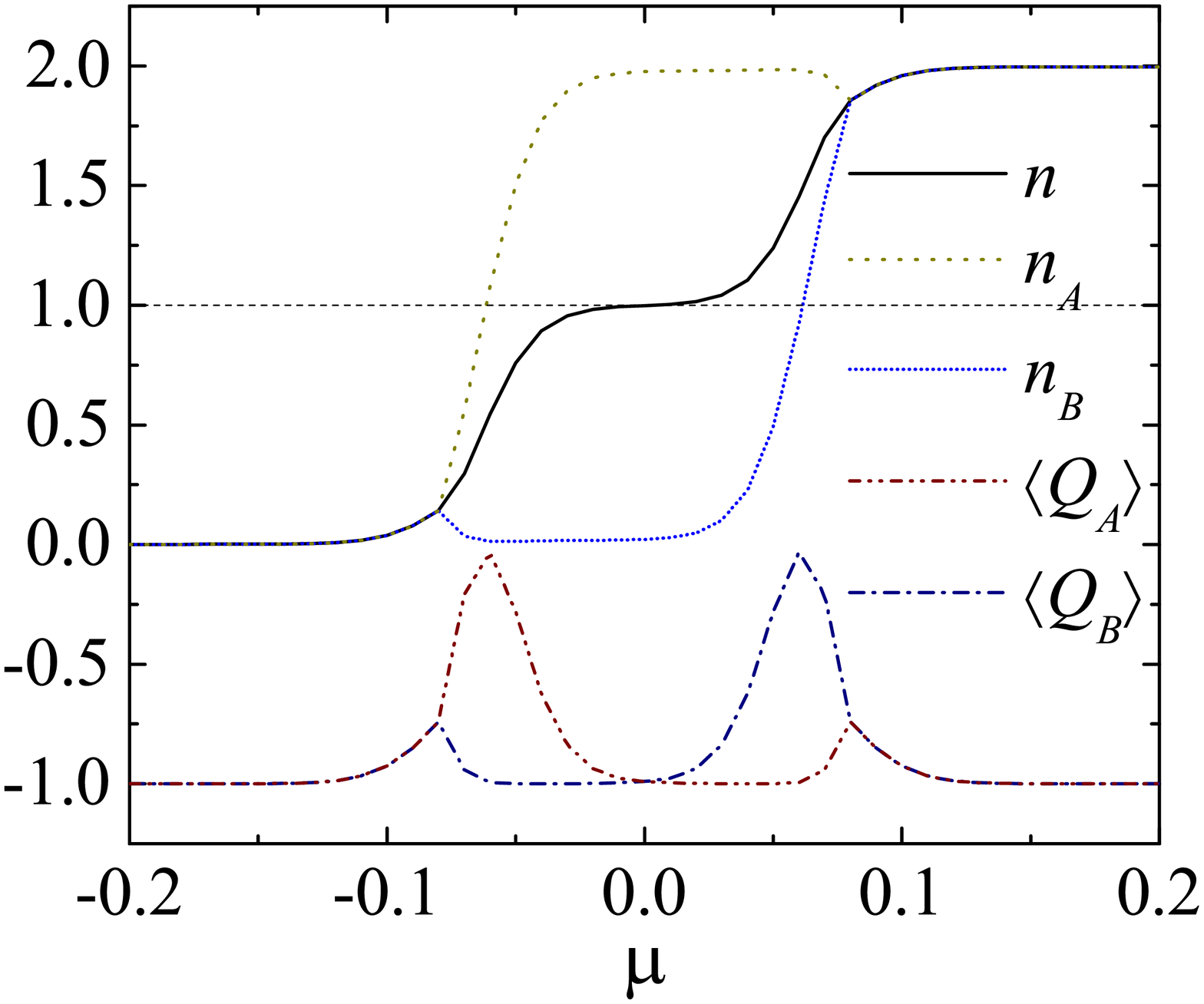}
\end{center}
\caption{(Color online) The total $n$ and the sublattice $n_A$, $n_B$ electron densities, $\langle Q_A\rangle$, $\langle Q_B\rangle$ via the chemical potential
for the interaction $U=8$ and temperature $T=0.01$. The dotted line shows the electron density $n=1$.}
\label{fig8}
\end{figure}

Similar to the HM phase, the CO phase is also separated into two distinct weak and strong correlation regimes. In Figs.~\ref{fig3} and \ref{fig4} we have already plotted the charge compressibility and the AF susceptibility in the CO phase. The CO phase occurs in the region $T<T_c$. One can see that both the charge compressibility and the spin susceptibility exhibit distinct behaviors in the weak and strong correlation regimes. In the weak correlation regime they are strongly suppressed like the ones in an insulator. However, in the strong correlation regime both the charge compressibility and the spin susceptibility are finite. This indicates the charge and spin excitations look like the ones in a metal.
The finite charge compressibility in the strong correlation regime can also be seen from the dependence of the electron density $n$ on the chemical potential. In Fig.~\ref{fig8} we plot the dependencies of the total and the sublattice electron densities, as well as $\langle Q_A\rangle$, $\langle Q_B\rangle$ as functions of the chemical potential in the strong correlation regime. It shows that the total electron density $n$ monotonously increases with the chemical potential. As a consequence, the charge compressibility is finite. One can also notice only at half filling
$\langle Q_A\rangle = \langle Q_B\rangle$. Away from half filling $\langle Q_A\rangle \neq \langle Q_B\rangle$, i.e the Ising variable is antiferromagnetically ordered. Any small shift of the chemical potential from its value at half filling drives the ordering of the Ising variable from homogeneous to staggered ones. The gap opening in the single-particle DOS does not generate a plateau in the function $n(\mu)$ around half filling $\mu=0$. The opposite behaviors of the single-particle excitation and its charge and spin counterparts show the CO phase is also fractionalized, like the ones in the HM phase. In the present model, the electron fractionalization occurs both at high and low temperatures. It smoothly crosses from the weak to the strong correlation regimes.
At a fixed temperature, the phase transition from the HM to CO states occurs before this crossover region from the normal to the fractionalized states when the interaction $U$ increases. The fractionalization appears only when the single-particle spectra open a gap and the correlations are not weak. Weak electron correlations cannot drive the system to the fractionalized state, since at high temperature they cannot open a gap in the single-particle spectra. Although at low temperature weak electron correlations can open a gap in the single-particle spectra due to the long-range ordering, they still cannot drive the system to the fractionalized state,  because in the weak correlation regime the charge compressibility and the spin susceptibility are suppressed like in an insulator, as they are shown in Figs. \ref{fig3} and \ref{fig4}.
The data plotted in Figs. \ref{fig3} and \ref{fig4} suggest in the strong correlation regime the charge compressibility and the spin susceptibility obey a universal scaling
\begin{eqnarray}
\frac{\chi}{U/t^*} &=& g_{\chi}(T/T_c),
\end{eqnarray}
where $\chi=\kappa$, $\chi_{\text{AF}}$, and
the scaling function $g_{\chi}(T/T_c)$ is independent on $U$. The scaling function $g_{\chi}(x)$ can empirically be fitted with the following function
\begin{equation}
g(x)= \frac{a}{x} \frac{\exp(b/x)}{[\exp(b/x) + c]^2},
\label{fitting}
\end{equation}
where $a$, $b$, $c$ are the fitting parameters. In Figs.~\ref{fig3} and \ref{fig4} we also plot the fitting function for a comparison. Although we cannot analytically derive the fitting function in Eq.~(\ref{fitting}), it fits well with the numerical results of the charge compressibility and the spin susceptibility in the CO phase.

\section{Conclusion}
We have showed the electron fractionalization in the symmetric three-component FKM. It is characterized by opposite behaviours of the single particles and their charge and spin counterparts. In the electron fractionalization the single particle spectra open a gap, while the charge and spin excitations are gapless. It occurs in both high- and low-temperature phases. When the interaction increases the ground state continuously changes from the normal state to the fractionalized one. At high temperature the phase is disordered, and strong electron correlations open a gap in the single-particle spectra, while the charge compressibility and the spin susceptibility remain finite like the ones in a metal. At low temperature the gap opening is due to a long-range ordering. In the strong correlation regime, despite the gap opening, the charge compressibility and the spin susceptibility are finite. They are universal functions of temperature in the fractionalized state.

So far we have only studied the special symmetric case of the three-component FKM.
The fractionalized state is unique at the special symmetric point in the surrounding phases of other natures.
It seems that the three-component FKM contains very rich physics, which has not fully been explored yet. The three-component FKM can also been considered as an extreme of the mass imbalance in the three-component Hubbard model \cite{Tran3}. The electron fractionalization in the three-component FKM suggests a possible fractionalization driven by the mass imbalance in the three-component Hubbard model. We leave this problem for further studies.

\section*{Acknowledgement}

This research is funded by Vietnam National Foundation
for Science and Technology Development (NAFOSTED) under Grant No 103.01-2017.13.


\begin{thebibliography}{99}

\bibitem{Giamarchi}
T. Giamarchi, {\em Quantum Physics in One Dimension}
(Clarendon Press, Oxford, 2004).

\bibitem{Senthil}
R. Nandkishore, M. A. Metlitski, and T. Senthil, Phys. Rev. B \textbf{86}, 045128 (2012).

\bibitem{Ng1}
Y. Zhou and T.-K. Ng, Phys. Rev. B \textbf{88}, 165130 (2013).

\bibitem{Ng2}
Y. Zhou, K. Kanoda, and T.-K. Ng,
Rev. Mod. Phys. \textbf{89}, 025003 (2017).

\bibitem{Schofield}
A. J. Schofield, Contemp. Phys. \textbf{40}, 95 (1999).

\bibitem{Assaad}
M. Hohenadler and F. F. Assaad, Phys. Rev. Lett. \textbf{121}, 086601 (2018).

\bibitem{Tran1}
D.-B. Nguyen and M.-T. Tran,
Phys. Rev. B \textbf{87}, 045125 (2013).

\bibitem{Tran2}
D.-A. Le and M.-T. Tran, Phys. Rev. B \textbf{91}, 195144 (2015).

\bibitem{Tran3}
D.-B. Nguyen, D.-K. Phung, V.-N. Phan, and M.-T. Tran, Phys.
Rev. B \textbf{91}, 115140 (2015).

\bibitem{Hubbard1}
J. Hubbard, Proc. R. Soc. (London) A \textbf{276}, 238 (1963).

\bibitem{Hubbard2}
J. Hubbard, Proc. R. Soc. (London) A \textbf{281}, 401 (1964).

\bibitem{Gutzwiller}
M. C. Gutzwiller, Phys. Rev. Lett. \textbf{10}, 159 (1963).

\bibitem{FK}
L. M. Falicov and J. C. Kimball, Phys. Rev. Lett. \textbf{22}, 997 (1969).


\bibitem{Lieb}
T. Kennedy and E. H. Lieb, Physica \textbf{138}A, 320 (1986).


\bibitem{Gruber}
C. Gruber and N. Macris, Helv. Phys. Acta \textbf{69}, 850 (1996).


\bibitem{Antipov}
A. E. Antipov, Y. Javanmard, P. Ribeiro, and S. Kirchner,
Phys. Rev. Lett. \textbf{117}, 146601 (2016).

\bibitem{Freericks}
J. K. Freericks, E. H. Lieb, and D. Ueltschi,
Phys. Rev. Lett. \textbf{88}, 106401 (2002).

\bibitem{Lemanski}
R. Lemanski, J. K. Freericks, and G. Banach,
Phys. Rev. Lett. \textbf{89}, 196403 (2002).

\bibitem{Kitaev}
 A. Kitaev, Ann. Phys. (Amsterdam) \textbf{321}, 2 (2006).

\bibitem{Fu}
W. Fu, Y. Gu, S. Sachdev, and G. Tarnopolsky, Phys. Rev. B \textbf{98}, 075150 (2018).

\bibitem{Metzner}
W. Metzner and D. Vollhardt, Phys. Rev. Lett. \textbf{62}, 324 (1989).

\bibitem{GKKR}
A. Georges, G. Kotliar, W. Krauth, and M. J. Rozenberg,
Rev. Mod. Phys. \textbf{68}, 13 (1996).

\bibitem{Mielsch1}
U. Brandt and C. Mielsch, Z. Phys. B \textbf{75}, 365 (1989).

\bibitem{Mielsch2}
U. Brandt and C. Mielsch, Z. Phys. B \textbf{79}, 295 (1990).

\bibitem{Mielsch3}
U. Brandt and C. Mielsch, Z. Phys. B \textbf{82}, 37 (1991).

\bibitem{FZ}
J. K. Freericks and V. Zlatic, Rev. Mod. Phys. \textbf{75}, 1333 (2003).

\bibitem{Sherson}
J. F. Sherson, C. Weitenberg, M. Endres, M. Cheneau, I. Bloch, and S. Kuhr,
Nature \textbf{467}, 68 (2010).

\bibitem{Bakr}
W. S. Bakr, A. Peng, M. E. Tai, R. Ma, J. Simon, J. I. Gillen, S. F\"olling,
L. Pollet, and M. Greiner, Science \textbf{329}, 547 (2010).

\bibitem{Greif}
D. Greif, M. F. Parsons, A. Mazurenko, C. S. Chiu, S. Blatt,
F. Huber, G. Ji, and M. Greiner, Science \textbf{351}, 953 (2016).

\bibitem{Brandt}
U. Brandt and M. P. Urbanek, Z. Phys. B \textbf{89}, 297 (1992).

\bibitem{Jordens} R. J\"{o}rdens, N. Strohmaier, K. G\"{u}nter, H. Moritz, and
T. Esslinger, Nature (London) \textbf{455}, 204 (2008).

\bibitem{Schneider} U. Schneider, L. Hackerm\"{u}ller, S. Will,
Th. Best, I. Bloch, T. A. Costi,
R. W. Helmes, D. Rasch, and A. Rosch, Science
\textbf{322}, 1520 (2008).

\bibitem{Yamazaki}
R. Yamazaki, S. Taie, S. Sugawa, and Y. Takahashi, Phys. Rev. Lett.
\textbf{105}, 050405 (2010).

\bibitem{Koga}
A. Koga, T. Saitou, and A. Yamamoto, J. Phys. Soc. Jpn. \textbf{82}, 024401 (2013).

\bibitem{Will}
S. Will, T. Best, U. Schneider, L. Hackerm\"uller, D.-S. L\"uhmann, and I. Bloch,
Nature (London) \textbf{465}, 197 (2010).

\bibitem{Zubarev}
D. N. Zubarev, Sov. Phys. Usp. \textbf{3}, 320 (1960).

\bibitem{Krauth}
M. Caffarel and W. Krauth, Phys. Rev. Lett. \textbf{72}, 1545 (1994).

\bibitem{NR}
W. H. Press, S. A. Teukolsky, W. T. Vetterling, and B. P. Flannery,
{\it Numerical Recipes in Fortran. The Art of Scientific Computing}, 2nd Edition (
Cambridge University Press, Cambridge, 1992).

\bibitem{Toschi}
A Toschi, P Barone, M Capone, and C Castellani, New J. Phys. \textbf{7}, 7 (2005).

\bibitem{Keller}
M. Keller, W. Metzner, and U. Schollw\"ock, Phys. Rev. Lett. \textbf{86}, 4612 (2001).

\bibitem{Georges}
A. Georges and G. Kotliar, Phys. Rev. B \textbf{45}, 6479 (1992).

\bibitem{Rozenberg}
M. J. Rozenberg, G. Kotliar, and X. Y. Zhang, Phys. Rev. B
\textbf{49}, 10181 (1994).




\end{thebibliography}
\end{document}